\documentstyle[12pt]{article}

\newcommand{\myref}[1]{(\ref{#1})}

\newcommand{\be}{\begin{equation}}
\newcommand{\ee}{\end{equation}}
\newcommand{\bea}{\begin{eqnarray}}
\newcommand{\eea}{\end{eqnarray}}
\newcommand{\T}{\theta}
\newcommand{\myskip}{\;\;\;\;\;\;\;\;\;\;}
\newcommand{\nn}{\nonumber}
\newcommand{\q}{i\pi}

\begin{document}

\titlepage

\hfill KFKI-1994-10/A

\hfill June 1994

\vskip 2truecm
\centerline{\bf POLYNOMIAL FORM FACTORS IN THE $O(3)$ NONLINEAR
$\sigma$-MODEL}
\vskip 1.5truecm
\centerline{{\bf J. Balog} and {\bf T. Hauer}}
\vskip 1.5truecm
\centerline{\it Research Institute for Particle and Nuclear
Physics} \smallskip
\centerline{\it H-1525 Budapest 114, P. O. B. 49, Hungary}
\vskip 2.5truecm
\centerline{\bf ABSTRACT}
\vskip 1.0truecm

We study the general structure of Smirnov's axioms on form factors
of local operators in integrable models. We find various
consistency conditions that the form factor functions have to
satisfy. For the special case of the $O(3)$ $\sigma$-model we
construct simple polynomial solutions for the operators of the
spin-field, current, energy-momentum tensor and topological
charge density.

\endtitlepage

In integrable quantum field theories physical quantities
can be determined
exactly in the bootstrap approach.
Making general assumptions about the properties of the S-matrix
has led to the complete determination
of the factorized scattering matrix of many well-known models
\cite{S1}-\cite{S4}.
The bootstrap method has also been applied to the determination
of matrix elements of local operators
in these theories. This form factor bootstrap was initiated in
\cite{FF1-1}-\cite{FF1-2} and further developed in refs.
\cite{FF2-1}-\cite{FF2-4}.
In addition to the explicit solutions for the Sine-Gordon, $SU(2)$
Gross-Neveu and $O(3)$ $\sigma$-models, the authors  of refs.
\cite{FF2-1}-\cite{FF2-4} also gave a complete set of axioms
that the form factors of an integrable model have to satisfy.
Using these axioms (which we will call the Smirnov axioms),
the form factor functions in other integrable models could
also be calculated \cite{FF3-1}-\cite{FF3-3}.

In this letter we study the general structure of Smirnov's axioms
and recast them into a form that is very convenient for the
recursive determination of the form factor functions in the $O(3)$
NLS model. We shall show that apart from a simple multiplicative
factor, which is explicitly given, the form factor functions in
the $O(3)$ model are polynomials in the rapidity variables.
These polynomials are determined by a simple recursive formula.
Although we do not claim originality on the solution of the form
factor functions we think that our approach will turn out to be
useful for the practical study of the structure of
operators in the $O(3)$ NLS model. (Note that while the solution
for the form factor functions of the $O(3)$ NLS model is
completely explicit in refs. \cite{FF2-3}-\cite{FF2-4}, the
complicated definitions
make them almost impossible to use in practical calculations.)

For simplicity, we consider integrable models describing
charge-selfconjugate bosons of equal, nonzero mass $m$ and no
bound states, as in the $O(3)$ NLS model. (In case of bound states
existing
the analytical properties of the form factors are more complicated and
model-dependent, and the existence of antiparticles makes the
formulae hard to
read because of upper and lower indices and charge conjugation matrices. We
think, nevertheless, that it is possible to generalize the
following arguments also for these cases.)

We parametrize the asymptotic states by the rapidities and the
internal indices
of the particles. These states are created from the physical vacuum by the
Zamolodchikov-Fadeev operators, whose exchange relation is
governed by the (two-particle) S-matrix:
\be
Z^{+}_{A}(\T)Z^{+}_{B}(\T')=S_{AB,YX}(\T-\T')
Z^{+}_{X}(\T')Z^{+}_{Y}(\T) \,.
\ee
Here the capital indices belong to the internal degrees of
freedom.

First we recall the list of requirements that in the bootstrap
approach the two-particle S-matrix is assumed to satisfy.
First,
it is meromorphic in the complex $\T$-plane and analytic in
the $0\leq \Im \T \leq \pi$ physical strip (no bound states).
Bose-symmetry and T-reversal invariance imply:
\bea
S_{AB,CD}(\T)&=&S_{BA,DC}(\T)\,,
\label{eq:S1}
 \\
S_{AB,CD}(\T)&=&S_{CD,AB}(\T)\,.
\label{eq:S2}
\eea
The most important assumptions in the S-matrix bootstrap are
unitarity,
crossing symmetry and Yang-Baxter equation, respectively:
\bea
S_{AB,CD}(\T)S_{CD,A'B'}(-\T) &=& \delta_{AA'}\delta_{BB'}\,,
\label{eq:S3}
 \\
S_{AB,CD}(i\pi-\T) &=& S_{AD,CB}(\T)\,,
\label{eq:S4}
\eea
\bea
S_{AB,VW}(\T)S_{VC,A'Z}(\T+\T')S_{WZ,B'C'}(\T') & &
\label{eq:S5}
\\
=\,\,S_{BC,VW}(\T')S_{AW,ZC'}(\T+\T')S_{ZV,A'B'}(\T)\,.& & \nn
\eea

Finally, we require that all the eigenvalues of the S-matrix at
zero
momentum transfer are $-1$ (which is satisfied in most
integrable models, including NLS):
\be
S_{AB,CD}(0)=-\delta_{AD}\delta_{BC}\,.
\ee
Let us now take a local field operator $X$ of spin $s$.  We
restrict our attention to its
matrix elements between the vacuum and an $n$-particle state.
(Other matrix
elements can be derived from these ones \cite{FF2-4}.) The
definition of the $n$-particle form factor is:
\be
\langle 0\vert X(0)\vert\T_1A_1,\ldots,\T_nA_n\rangle =
f^{(n)}_{A_1\ldots A_n}(\T_1,\ldots,\T_n)\,.
\ee

The form factors are originally defined for ordered sets of real
rapidities
corresponding to the asymptotic states, but they can
be analytically continued to
the complex plane in all variables. The Smirnov axioms postulate
the properties of this analytically extended $f$ function:
\bea
\label{eq:spin}
f^{(n)}_{A_1\ldots A_n}(\T_1,\ldots,\T_n) &=& e^{\lambda s}
f^{(n)}_{A_1\ldots A_n}(\T_1-\lambda,\ldots,\T_n-\lambda)\,, \\
\label{eq:transp}
f^{(n)}_{\ldots AB\ldots}(\ldots,\T,\T',\ldots) &=&
S_{AB,YX}(\T-\T') f^{(n)}_{\ldots
XY\ldots}(\ldots,\T',\T,\ldots)\,, \\
\label{eq:cyclic}
f^{(n)}_{A_1A_2\ldots A_n}(\T_1,\T_2\ldots,\T_n) &=&
f^{(n)}_{A_2\ldots A_nA_1}(\T_2,\ldots,\T_n,\T_1-2i\pi )\,,
\eea
\bea
{\rm Res}(f^{(n+2)}_{ABU_1\ldots
U_n}(\beta+i\pi,\beta,\T_1,\ldots,\T_n) =
\frac{i}{2\pi}\left( \delta_{AB}f^{(n)}_{U_1\ldots U_n}
(\T_1,\ldots,\T_n)\right.
 \nn \\
- \left.S_{BU_1\ldots U_n,V_1\ldots
V_nA}(\T_1,\ldots,\T_n\vert\beta)
f^{(n)}_{V_1\ldots V_n} (\T_1,\ldots,\T_n)\right)\,.
\label{eq:resid}
\eea
The matrix $S(\T_1,\ldots,\T_n\vert\beta)$ entering
\myref{eq:resid}
is a product of two-particle S-matrices corresponding to the scattering
of particles $(\beta B,\T_1U_1,\ldots,\T_nU_n)$ into the set
$(\T_1V_1,\ldots,\T_nV_n,\beta A)$. One should also postulate that the
functions $f$ are meromorphic in all variables and they are
analytic in the physical strip except for poles explicitly given
in
\myref{eq:resid}. (In case of bound states giving extra
(nonkinematical)
poles a fifth equation applies connecting the $(n+1)$-particle
form factor to the $n$-particle one.)

We can simplify eq. \myref{eq:spin} if we write
$f$ as a product of a ``scalar" form factor $F$ and an overall
factor that carries the Lorentz-transformation character and
cancels the $e^{s\lambda}$ in eq. \myref{eq:spin}:
\be
\label{eq:spin2}
f^{(n)}_{A_1\ldots A_n}(\T_1,\ldots,\T_n) =
\left(\sum_{i=1}^ne^{\T_i}\right)^s
F^{(n)}_{A_1\ldots A_n}(\T_1,\ldots,\T_n)\,.
\ee

Equations \myref{eq:transp}, \myref{eq:cyclic} and
\myref{eq:resid} are of the same form (with $F$ instead of $f$),
except for \myref{eq:resid} in the two-particle case,
which changes to:
\be
(\T_1-\T_2-i\pi)^{s}
F_{A_1A_2}^{(2)}(\T_1,\T_2) = \hbox{finite}\,,\myskip
(\T_1 \rightarrow \T_2+i\pi)\,.
\ee

The first three axioms together describe an invariance property of
the form factor function $F$, while
\myref{eq:resid} governs the structure of the poles.
To identify the symmetry group described by the first three axioms, we
summarize the way it is realized.  Let us consider a group ${\cal G}$, a
manifold ${\cal M}$, and the set of functions mapping the manifold into a
linear space:  ${\cal F} = \{V:{\cal M}\rightarrow{\cal L}\}$.  Let $\varphi$
be an action of the group ${\cal G}$ on ${\cal M}$:
\be
\varphi(\varphi(p,g_1),g_2)=\varphi(p, g_1g_2)\,, \myskip \forall
g_1, g_2 \epsilon {\cal G}\,, \myskip p \epsilon \cal M\
\ee
and assume that for all $p\epsilon {\cal M}$, $g\epsilon {\cal G}$
there corresponds a
nonsingular linear operator $M_g(p)$ acting on ${\cal L}$
satisfying
\be
M_{g_1}(p)M_{g_2}(\varphi_{g_{1}}(p)) = M_{g_1g_2}(p)\,.
\label{eq:Mprop}
\ee

Now we can define an action of the group on the set $\cal F$ of
linear
functions, $\Phi:{\cal G}\times {\cal F} \rightarrow {\cal F}$
as:
\bea
[\Phi(g,V)](p) &\equiv& M_g(p)V(\varphi(p,g))\,, \\
\Phi(g_1,\Phi(g_2,V))& =& \Phi(g_1g_2,V)\,, \myskip \forall
g_1, g_2 \epsilon {\cal G}\,, \myskip p \epsilon \cal M \,.
\eea

Introducing a compact notation for the action of $\varphi$ and
$\Phi$, this can be summarized as:
\bea
\varphi(p,g) &\equiv& p^g\,, \\
\Phi(g,V) &\equiv& _gV\,, \\
M_{g_1}(p)M_{g_2}(p^{g_1}) &=& M_{g_1g_2}(p)\,, \\
\ _{g}V(p) &=& M_g(p)V(p^g)\,.
\eea

The invariance group of the form factor functions can now be identified
by
studying the action on the arguments. (The elements of the
manifold $\cal M$ are
$n$-component rapidity vectors.)  It is easy to see that the
transformations form
a direct product group with a continous and a discrete component.  The
continous component is the (1+1)-dimensional Lorentz group
$\bf L$ acting
through shifting all the rapidities by a common value  as in eq.
\myref{eq:spin}.
The discrete component contains all the permutations of the $n$ variables
generated by the transpositions \myref{eq:transp}. In addition,
any of the $n$
rapidities can be shifted by $2ki\pi$, where $k$ is an integer.
(One can construct these
transformations by repeatedly using \myref{eq:transp} and
\myref{eq:cyclic}.)
Since the permutations act on the generators of this latter group
as well, the discrete component is a half-direct product. (Note
that this
discrete group can be generated by the two elements corresponding
to \myref{eq:transp} and \myref{eq:cyclic}.) Thus the symmetry
group of the first three axioms is:
\be
{\cal G} = {\bf L}\otimes({\bf S}_n\wedge{\bf Z}^{\otimes n})\,.
\ee

It is easy to find the matrices corresponding to the linear
transformations in eqs. \myref{eq:spin}-\myref{eq:cyclic}
explicitly and verify that as a consequence of the
S-matrix properties \myref{eq:S1}-\myref{eq:S5}
they satisfy the consistency
relation \myref{eq:Mprop}. This means that
the first three axioms can be summarized as:
\be
_{\cal G}F = F\,.
\ee

Now we turn to the fourth Smirnov axiom which determines the
singularity structure of the form factors.
We write the form factor $F$ as a product of two terms, an
overall
factor which carries all the singularities and a regular term with the
tensorial structure of $F$:
\be
F^{(n)} = \left(\frac{\pi}{2}\right)^{n-1}
 \left(\prod_{i<j}\Psi(\T_i-\T_j)\right)G^{(n)}\,.
\ee
Here the function $\Psi$ is chosen to satisfy
\bea
{\rm Res}(\Psi(z);z=i\pi) &=& -\frac{4}{\pi^2}\,, \\
\Psi(i\pi+\T)&=&-\Psi(i\pi-\T)\,.
\eea

The advantage of using the reduced form factor $G^{(n)}$
is that while it is still invariant under the group $\cal G$,
it is analytic in the physical strip and the fourth axiom
directly
gives its value at the point $\T_1-\T_2=\q$.
More precisely, if we define
\bea
E(\T) &\equiv& \frac{1}{\Psi(\T)\Psi(\T+i\pi)}\,, \\
R_{AB,CD}(\T) &\equiv& -S_{AB,CD}(\T)E(\T)\,,\\
{\tilde S}_{AB,CD}(\T) &\equiv& R_{AB,CD}(\T)/E(-\T)\,,
\eea
then $\tilde S$ plays the role of the ``reduced" S-matrix and
$G$  satisfies \myref{eq:transp} with $\tilde S$ instead of
$S$ and \myref{eq:cyclic} with an extra $(-1)^{n-1}$ factor
inserted at the rhs. These modified equations, together with
\myref{eq:spin}, define transformation matrices  also satisfying
the relations \myref{eq:Mprop} and they can be compactly
written as
\be
^{\cal G}G = G\,,
\label{eq:red13}
\ee
where the upper left index notation now indicates this
new, modified group action. In the rest of the paper we will
use this modified action only.

\myref{eq:resid} now reduces to
\bea
\label{eq:red4}
G^{(n)}_{ABU_3\ldots U_n}(\beta+i\pi,\beta,\T_3,\ldots,\T_n) &= &
T_{ABU_3\ldots U_n,V_3\ldots V_n}(\beta\vert\T_3,\ldots,\T_n) \nn\\
&\times& G^{(n-2)}_{V_3\ldots V_n}(\T_3,\ldots,\T_n)
\eea
except for the $n=2$ case which is modified to

\be
(\T_1-\T_2-i\pi)^{\vert s\vert-1} G_{A_1A_2}^{(2)}(\T_1,\T_2) =
\hbox{finite,}\myskip (\T_1 \rightarrow \T_2+i\pi)\,.
\label{eq:red4b}
\ee

The matrix appearing in \myref{eq:red4} can be explicitly given in
terms of the matrix $R$ and the function $E$:
\bea
&T_{ABU_3\ldots U_n,V_3\ldots V_n}(\beta\vert\T_3,\ldots,\T_n)
=&\nn\\
&\myskip\frac{1}{2i\pi}( E(\beta-\T_3)\ldots E(\beta-\T_n)
\delta_{AB}\delta_{U_3V_3}\ldots\delta_{U_nV_n}  &\\
&\myskip + (-1)^{n-1}R_{BU_3,X_4V_3}(\beta-\T_3)\ldots
R_{X_nU_n,AV_n}(\beta-\T_n))\,.&\nn
\eea

Note that eq. \myref{eq:red4} gives the value of the $n$-particle
reduced
form factor in terms of the $(n-2)$-particle one at the specified
point (and not
its residue as \myref{eq:resid}). Now (after choosing a suitable
function $\Psi$) our task is to determine a series of {\em analytic} functions
$G^{(n)}$ that solves eqs. \myref{eq:red13} - \myref{eq:red4b}.

If we know the $(n-2)$-particle reduced form factor, eq.
\myref{eq:red4} gives
the value of the $n$-particle one at the special point. Assuming
that the $n$-particle
form factor also solves the equations, one can apply the elements
of the group ${\cal G}$ and compute the values of this
function at other special points
as well. We can compute the
value
of $G^{(n)}$ at all points where any two of its arguments differ
by an odd
multiple of $i\pi$.
This opens the possibility of determining the form factors
completely. If the form factors belong to some special class
of functions, the knowledge of their values at some specially
chosen points could be sufficient for their determination.
This is the case for the $O(3)$ NLS model form factors, since
the reduced form factors in this model, as we shall see,
are polynomials.

However, a question of consistency arises here.  It is easy to
see that the
determination of the value of $G^{(n)}$ at a special point
$(\T_1,\ldots,\T_n)$ defined above is not neccessarily
well-defined, since in general there are many
different points $(\T'_2+i\pi,\T'_2,\ldots,\T'_n)$ from which one can get there
by  applying the elements of ${\cal G}$. If the system of the equations is
consistent,
the different ways determining $G^{(n)}$ at a given
special point should
give the same result. This condition gives consistency equations.

Now we present
some equations of this kind, which will play an important role in
the study of the $O(3)$-model form factors.
To make the formulae easier to read we introduce a compact notation: dot over
(under) a variable means a shift by $i\pi$ ($-i\pi$), and we denote the $k$th
$\T_k$ rapidity simply by its index $k$. We define the following functions
built from the $(n-2)$-particle form factor:
\bea
H^{(2)}_{ABU_3\ldots U_n}(12\ldots n) &=& T_{ABU_3\ldots
U_n,V_1\ldots V_3} (2\vert 3\ldots n)\nn \\
&&\myskip \times G^{(n-2)}_{V_3\ldots V_n}(3\ldots n)\,,
\label{eq:H2} \\
H^{(k)} & = & ^{P_k\ldots P_3}H^{(2)}\,, \qquad k=3,\dots,n\,.
\label{eq:Hk}
\eea

Here the $P_i$s are those group elements that represent the
transposition of the
$(i-1)$th and the $i$th component of the rapidity vector.
The $H^{(k)}$s are nothing but the values of $G^{(n)}$ at
special points:
\be
G^{(n)}(\dot{k}23\ldots n) = H^{(k)}(12\ldots n)\,.
\label{eq:iden}
\ee

We will use the following consistency equations satisfied
by the $H^{(k)}$s:
\bea
H^{(k)}(\ldots\T_k\ldots\T_l\ldots) & = &
H^{(l)}(\ldots\T_k\ldots\T_l\ldots)\,, \myskip \T_k = \T_l\,,
\label{eq:konz1}
\\
^{P_{s}}H^{(k)} & = & H^{(k)}\,, \myskip s \not= k,k+1\,,
\label{eq:konz2}
\\
^{P_{k}}H^{(k)} & = & H^{(k-1)}\,,
\label{eq:konz3}
\\
^{P_{k+1}}H^{(k)} & = & H^{(k+1)}\,,
\label{eq:konz4}
\\
\label{eq:konz5}
H^{(2)}_{A_1\ldots A_n}(12\ldots n) & = &
(-1)^{n-1}H^{(n)}_{A_2\ldots A_nA_1}(13\ldots n{\d 2})\,,
\\
H^{(k)}_{A_1\ldots A_n}(1\dot{l}3\ldots n) & = &
(-1)^{n-1}H^{(l-1)}_{A_2\ldots A_nA_1}(13\ldots n
{\mbox{\d{\it k}}})\,.
\label{eq:konz6}
\eea

Using the identification \myref{eq:iden} and assuming that
$G^{(n)}$
satisfies the Smirnov axioms it is easy to understand the meaning
of these consistency equations. However, what we need is to
prove them directly by using the definitions \myref{eq:H2},
\myref{eq:Hk} and the fact that the $(n-2)$-particle
form factors $G^{(n-2)}$ entering these definitions satisfy the
Smirnov axioms. Indeed,
one can show that if the two-particle S-matrix possesses all the
properties \myref{eq:S1}-\myref{eq:S5}
and all the
$(n-2)$ and
the $(n-4)$-particle reduced form factors (from which the $H$'s
are generated) satisfy the four Smirnov's axioms, then
\myref{eq:konz1}-\myref{eq:konz6} are satisfied.

Before turning to the special case of the $O(3)$ model we note
that the Smirnov axioms alone cannot determine the form factors
completely. Indeed,
it is easy to see that from a set of solutions $G^{(n)}$ we can
generate new ones if we
multiply the form factors by  scalar functions which are ${\cal
G}$-symmetric
\be
\Omega^{(n)}\left((\T_1,\ldots,\T_n)^g\right) =
\Omega^{(n)}(\T_1,\ldots,\T_n)\,,\myskip \forall g\epsilon
{\cal G} \ee
and satisfy
\be
\Omega^{(n+2)}(\beta+i\pi,\beta,\T_1,\ldots,\T_n) =
\Omega^{(n)}(\T_1,\ldots,\T_n)\,.
\ee
An important example of such functions is provided by the
invariant squared mass of the $n$-particle state:
\be
\mu^2(\T_1,\ldots,\T_n)  \equiv
\left(\sum\cosh(\T_i)\right)^2 -
\left(\sum\sinh(\T_i)\right)^2 \,.
\ee

{}From now on we consider the $O(3)$ $\sigma$-model, where our
considerations lead to the complete determination of the
form factor functions.
The S-matrix of the model is given by \cite{S2}:
\be
S_{AB;CD}(\T)=
S_1(\T)\delta_{AB}\delta_{CD}+S_2(\T)\delta_{AC}\delta_{BD}+
S_3(\T)\delta_{AD}\delta_{BC}\,,
\ee
where
\bea
S_1(\T)&=&\frac{2\q\T}{(\T+\q)(\T-2\q)}\,,
\nn\\
S_2(\T)&=&\frac{\T(\T-\q)}{(\T+\q)(\T-2\q)}\,,
\label{eq:O3S}
\\
S_3(\T)&=&\frac{2\q(\q-\T)}{(\T+\q)(\T-2\q)} \nn
\eea
and we choose the well-known two-particle form factor for $\Psi$:
\bea
\Psi(\T)&=&\frac{\T-\q}{\T(2\q-\T)}\tanh^2\frac{\T}{2}
\,,
\\
E(\T)&=&(\T+\q)(\T-2\q)\,.
\eea

The fact that makes the form factors explicitly calculable in this
case is that
the function $E$ is identical to the denominator of the S-matrix
\myref{eq:O3S}.
{}From this it follows that the $R$-matrix is ($-1\times$) the
numerator of $S$, thus it is a polynomial of $\T$.  Since both $E$ and $R$ are
polynomials, the matrix $T$ entering \myref{eq:red4} is also a
polynomial of the
rapidities.  One can easily check that its degree is $(2n-5)$ in
$\beta$ and that it is
quadratic in the other rapidities.  From this it follows that if $G^{(n-2)}$
is a polynomial, the function $H^{(2)}$ is also a
polynomial and it is
straightforward to prove that all the $H^{(k)}$s are
polynomials.

More precisely the following statement holds:  if $G^{(n-2)}$ satisfies the
axioms and it is a polynomial in all its variables of maximum degree
$(n-4)$,
then the $H^{(k)}$s are polynomials of maximum degree
$(2n-5)$ in $\T_k$ and $(n-2)$
in the other arguments.

Now it is a natural assumption that $G^{(n)}$ is
also a polynomial and we can write the following Ansatz:
\be
G^{(n)}(12\ldots n) = \sum_{k=2}^nH^{(k)}(12\ldots n)
\prod_{\scriptstyle l=2 \atop \scriptstyle l\not=k}^n
\frac{(1\dot{l})}{(kl)}\,,
\label{eq:ans}
\ee
where
\be
(kl)\equiv (\T_k-\T_l)\,.
\ee

\myref{eq:ans} is the main result of this paper. Together with the
definitions \myref{eq:H2} and \myref{eq:Hk} it provides a
recursive polynomial solution for all the form factors of the
model. In writing down \myref{eq:ans} we have made use of the
fact that a polynomial of degree $(n-2)$ (which we assume
$G^{(n)}$ is in the variable $\T_1$) is determined by its values
at $(n-1)$ different points. This is precisely the information
contained in \myref{eq:iden}.

It can be proven that \myref{eq:ans} indeed defines a
polynomial solution of the Smirnov axioms. First,
one can show that if the $(n-2)$-particle form factor
satisfies Smirnov's
axioms and it is a polynomial of maximal degree $(n-4)$ in all
variables, then the
function $G^{(n)}$ constructed by \myref{eq:ans} is a polynomial
in all its
variables of maximal degree $(n-2)$. Here one has to use
\myref{eq:konz1} and
the counting of the degrees is straightforward.
One can then check that eq. \myref{eq:transp} is satisfied by
\myref{eq:ans}
by verifying it at $(n-1)$ points (which is sufficient since
both sides are polynomials of degree $(n-2)$). Here one
uses \myref{eq:konz2}-\myref{eq:konz4}. \myref{eq:cyclic} is also
satisfied at these points as a consequence of \myref{eq:konz5} and
\myref{eq:konz6}. Finally, the last Smirnov axiom is satisfied by
\myref{eq:ans} by construction.

To summarize, \myref{eq:ans} provides us with a whole family of
form factor functions corresponding to the matrix elements of the
local field operator between the vacuum and an increasing number
of particles, provided a starting element of
this family (corresponding to $n$ particles) is known,
it is a polynomial satisfying Smirnov's axioms and its degree is
not higher than $(n-2)$.

Studying the one- and two-particle form factors of the model, we
have
found that the four basic operators (spin-field, current,
energy-momenum tensor,
topological charge) all have matrix elements of the type described above.

We define the reduced form factors of these four operators,
respectively, as:
\bea
f^{a}_{A_1\ldots A_n}(1\ldots n) &=& \Psi_0
       G^{(spin)a}_{A_1\ldots A_n}(1\ldots n)\,,
\\
f^{\pm;a}_{A_1\ldots A_n}(1\ldots n) &=&
\left(\sum_{i=1}^ne^{\pm\T_i}\right) \Psi_0
       G^{(curr.)a}_{A_1\ldots A_n}(1\ldots n)\,,
\\
f^{\pm\pm}_{A_1\ldots A_n}(1\ldots n) &=&
\left(\sum_{i=1}^ne^{\pm\T_i}\right)
\left(\sum_{i=1}^ne^{\pm\T_i}\right) \Psi_0
       G^{(E-M)}_{A_1\ldots A_n}(1\ldots n)\,,
\\
f_{A_1\ldots A_n}(1\ldots n) &=&
(\mu^2-1)\Psi_0
       G^{(top.)}_{A_1\ldots A_n}(1\ldots n)\,,
\eea
where
\be
\Psi_0 =
\left(\frac{\pi}{2}\right)^{n-1} \prod_{i<j}\Psi(\T_i-\T_j)\,.
\ee

Note that the operators are defined by their Lorentz- and isospin
transformation character. In case of the Lorentz-vector and (1,1)-tensor the
conservation is included in the definition. The current and E-M tensor have
non-vanishing form factors for an even number of particles only,
whereas
the spin and topological charge operators have odd particle form
factors only.
Since there is no invariant isovector, the family of form factors
of the
topological charge begins with the three-particle case and that is
why the
$\mu^2-1$ factor has been introduced here to cancel the unwanted
poles of the three-particle matrix element.

The starting elements of the respective families are:
\bea
G^{(spin)a}_{A}(\T) &\equiv& \delta_{aA}\,,
\\
G^{(curr.)a}_{A_1A_2}(\T_1,\T_2) & \equiv & \epsilon_{aA_1A_2}\,,
\\
G^{(E-M)}_{A_1A_2}(\T_1,\T_2) &  \equiv &
\frac{1}{i\pi-(\T_1-\T_2)}\delta_{A_1A_2}\,,
\\
G^{(top.)}_{A_1A_2A_3}(\T_1,\T_2,\T_3) &\equiv & \epsilon_{A_1A_2A_3}
\,.
\eea

It is easy to see that these functions satisfy the axioms.  The
form
factors of the current and topological charge are really polynomials of
the right degree
and the recursion works automatically.  In the case of the spin and E-M
tensor this condition does not hold, but one can show that
the Ansatz \myref{eq:ans} does give the next element of the family
(i.e. the function we get
satisfies the axioms as well) and the three-particle spin and four-particle E-M
tensor form factors are polynomials of degree $1$ and $2$, respectively.
This means that we can use our recursive construction also for
these operators.

We think that the recursive formula \myref{eq:ans} will prove to
be
useful in the problem of studying the correlation functions in the
$O(3)$ NLS model. The problem of constructing the correlation
functions by summing over an infinite number of possible
intermediate states
is discussed in  refs.
\cite{FF3-1},\cite{FF3-3} for the case of integrable models
with no internal quantum numbers. Some preliminary results
in the case of the more difficult problem of the $O(3)$ model
are discussed in \cite{O3}.

\end{document}